\documentclass[12pt]{iopart}
\usepackage{iopams}
\usepackage{graphicx}
\usepackage{setspace}
\usepackage{booktabs}
\usepackage{cite}

\begin{document}

\title[A Portable Diagnostic Device for Cardiac Magnetic Field Mapping ]{A Portable Diagnostic Device for Cardiac Magnetic Field Mapping}

\author{J. W. Mooney, S. Ghasemi-Roudsari, E. Reade Banham, C. Symonds, N. Pawlowski, and B. T. H. Varcoe }

\address{School of Physics and Astronomy, University of Leeds, Leeds LS2 9JT, UK}
\vspace{10pt}
\begin{indented}
\item[]June 2016
\end{indented}

\begin{abstract}
	\noindent In this paper we present a portable magnetocardiography device.
	The focus of this development was delivering a rapid assessment of chest pain in an emergency department.
	The aim was therefore to produce an inexpensive device that could be rapidly deployed in a noisy unshielded ward environment.
	We found that induction coil magnetometers with a coil design optimized for magnetic field mapping possess sufficient sensitivity ($104fT/\sqrt{Hz}$ noise floor at 10Hz) and response ($813fT/\mu V$ at 10Hz) for cycle averaged magnetocardiography and are able to measure depolarisation signals in an unshielded environment. We were unable to observe repolarisation signals to a reasonable fidelity.
	We present the design of the induction coil sensor array and signal processing routine along with data demonstrating performance in a hospital environment.

\end{abstract}

%
%
%
%
%
\section{Introduction}
Chest pain is responsible for one of the highest rates of emergency hospital visits in industrialized countries~\cite{Sanchez2007} and accounts for a large proportion of hospital admissions.
Statistics show that around 75\% of patients who present at the Emergency Department with chest pain do not have a cardiac related condition~\cite{Six2012,Backus2013,Backus2011,Rohacek2012}, yet they still need to go through a full diagnostic pathway which can take more than 10 hours~\cite{Six2012}.
This leads to several thousand people occupying bed spaces placing an additional burden on health care systems. A diagnostic that is capable of rapidly stratifying the cases and removing those patients who don't need an overnight stay is therefore valuable in both triage and cost saving~\cite{Six2012}.

Magnetocardiography (MCG) involves capturing Magnetic Field Maps (MFM's) of current distributions resulting from cardiac action potentials~\cite{Eskola1987,Korhonen2000,Korhonen2002,Malmivuo1995,McFee1972,Moshage1996,Ramon1998,Saarinen1974,Smith2001}.
It has been shown that MCG gives significant improvements in diagnostic capability over an ECG~\cite{Agarwal2012,Fenici2013,Gapelyuk,Hailer2005,Korhonen2006,Leithauser2011,Lim2007,Lim2009,Park2005,Smith2006b,Steinisch2013,Tolstrup2006}.
Significantly, in this respect, it has been demonstrated that MCG is capable of reliable detection of Non-ST-Elevated Myocardial Infarction (NSTEMI)~\cite{Agarwal2012,Lim2009}, which are by definition difficult to detect using ECG\@.
For this reason all ECG negative chest pain patients are treated as having an NSTEMI until other diagnostic results can be obtained~\cite{Backus2013}.
Hence, the short time to produce a MCG (typically \textless10 minute measurement) dramatically reduces the time for diagnosis and removes otherwise healthy patients earlier in the process and is therefore a tool with obvious clinical benefits.

The principle focus of the current research was the creation of a portable MCG device that would be capable of providing a rapid assessment of acute coronary syndrome (ACS) in an Emergency Department. To meet this goal, the device requirements are sensitivity to a magnetic window of between 0.1pT and 300pT, in the frequency range of around 1--40 Hz~\cite{Bison2009} and a spatial resolution sufficient to detect anomalies with a spacing of 10--15cm (for a sensor operated 10cm from the chest wall)~\cite{Guofa2011}.

Cardiac MFM devices typically use an array of sensitive magnetometers detectors to collect the magnetic field of the heart by simultaneously sampling at many positions across the chest.
Sensors include liquid helium cooled SQUID detectors, which have been used in commercially available devices for over 40 years~\cite{Stroink2010}, and atomic physics detectors and giant magnetoresistance detectors have also been developed~\cite{Bison2009,Pannetier2011, Shah2013}.

These devices are not always suitable for an Emergency Department as the associated apparatus is bulky, they often require liquid helium, specialist training to use, they are fixed in place and typically require an electromagnetically shielded room.
In contrast, induction coil magnetometers have been used several times for cardiac magnetic field detection by several authors.
They meet the demands of signal sensitivity~\cite{Baule1970,Baule1963,Cohen1969,Cohen1967,Estola1982,Tashiro2006}, they are inexpensive, do not require cooling and can be run from batteries.
However earlier efforts required noisy high gain amplifiers, large and heavy coils which are unsuitable for magnetic field mapping and fixed electronically implemented gradiometer arrangements.
Here we present a more compact coil design which when combined with modern analog-to-digital converters (ADC's) and digital signal processing (DSP) produces a device capable of detecting the cardiac magnetic field with an array of 19 sensors.

We first present the design of the sensors, array and DSP routine. Then we show that this design has the capability of resolving the field of the heart within both shielded and unshielded environments.

\section{Apparatus}
\subsection{Coil Design}
The most important aspect of this device is the construction of the sensor elements to achieve both the sensitivity and required spatial resolution.
An induction coil magnetometer will have an output voltage determined by

\begin{equation}
\label{outputVoltage}
V=AN\frac{{\rm d}B(t)}{{\rm d}t}=ANB2\pi f
\end{equation}

where $N$ is the number of windings, $A$ is the effective cross sectional area of the coil, $B(t)$ is the time varying magnetic field, with a magnitude $B$, and $f$ is the frequency of oscillation of the field.
The smallest field, $S$, that can be detected given thermal Johnson noise resulting from the winding resistance is given by:
 detected given thermal Johnson noise resulting from the winding resistance is given by:

\begin{equation}
\label{sensitivity}
S=\frac{\sqrt{4k_{B}TR_{a}}}{2\pi fNA}
\end{equation}

where $k_{B}$ is Boltzmans constant, $T$ is the temperature, $R_a$ is the antenna wire resistance given by:

\begin{equation}
R_{a}=N2\pi^{2}a^{2} \rho r_{coil}
\end{equation}

where $a$ is the radius and $\rho$ is the resistivity of the wire used in the windings on a circular coil of average winding radius $r_{coil}$. Equations~\ref{outputVoltage},~\ref{sensitivity} can be used to find the coil structure with lowest noise level given the design constraints.

If the coil parameters are the length $L$, the coil outer diameter, $D$, and the coil inner diameter, $D_{i}$, the dimensions that give the lowest noise level have the ratio $D_{i}:D = 0.425:1$.
In addition to this we primarily want to measure the component of the magnetic field aligned to the axis of the coil.
Zilstra~\cite{Zijlstra1967} notes that the optimum coil structure to measure the axial component of the magnetic field is achieved when $L/D= 0.69$ for the above ratio $D_{i}/D$.
The coil diameter itself is determined according to the desired device resolution, leaving the radius of the wire only remaining free parameter in the coil design.

The output voltage of the coil is determined by N. As all other parameters are now fixed, voltage is determined exclusively by the wire radius $a$.
A thinner wire increases the voltage output at the expense of increased coil resistance and subsequently increased noise, leading to a fixed signal to noise ratio irrespective of wire diameter.

\Tref{tabone} presents outputs from a simulation of the coil design presented here (MFM Coil) compared to the Brooks coil of the same outside dimension and wire diameter, $a=0.23mm$. A Brooks coil is a special case in which the ratio of the dimensions are chosen to optimise inductance for which the ratio of the dimensions are $D:D_{i}:L=4:2:1$.
The current MFM Coil design has a higher voltage and a lower noise equivalent field at the target frequency of 30Hz than the Brooks coil.
The noise equivalent field is the smallest field strength that could be measured above Johnson noise of the detector.
The gains are a factor of 1.6 in signal to noise and a factor of about 2.9 in output voltage.

The increased signal size plays a role in the subsequent electronics especially when thermal and Johnson noise in the electronics is similar in size to the cardiac signal.
Overall both effects improve data collection times by a factor of 20.
This is important because, while the gains are modest, the overall impact on the design is a significant reduction (more than an order of magnitude) in the data collection time when cycle averaging is used.
\begin{table}
	\caption{\label{tabone}Table comparing the classic Brooks coil design with the MFM coil design presented in this paper.}
	\begin{indented}
		\item[]
		\begin{tabular}{llllcc}
			\toprule
			Coil & $D$ & $D_{i}$ & $L$ & $V_{out}$ at 1pT (40Hz) & Noise Equivalent Field \\
			\midrule
			MFM & 12cm  & 5.1cm & 8.28cm & 616nV & 57fT \\
			Brooks & 12cm  & 6cm & 3cm & 211nV & 96fT \\
			MFM & 8.5cm  & 3.6cm & 5.87cm & 155nV & 136fT \\
			Brooks & 8.5cm  & 4.25cm & 2.125cm & 53nV & 227fT \\
			MFM & 4.25cm  & 1.8cm & 2.9cm & 9.5nV & 773fT \\
			Brooks & 4.25cm  & 2.125cm & 1.0625cm & 3.3nV & 1.3pT \\
			\bottomrule
		\end{tabular}
	\end{indented}
\end{table}

\subsection{Mapping Array Construction}

The commercial analog to digital converter (ADC) we used has 16 channels. One channel was reserved for the ECG trigger, leaving 15 cardiac magnetometer channels.
MFM Coils with a diameter of 7cm were chosen to cover the measurement area of $\sim 25\times25 cm$ in a hexagonal array, arranged in order to detect the principle components of the hearts magnetic dipole field.

\begin{figure}[p]
	\centering
	\includegraphics[width=0.8\textwidth]{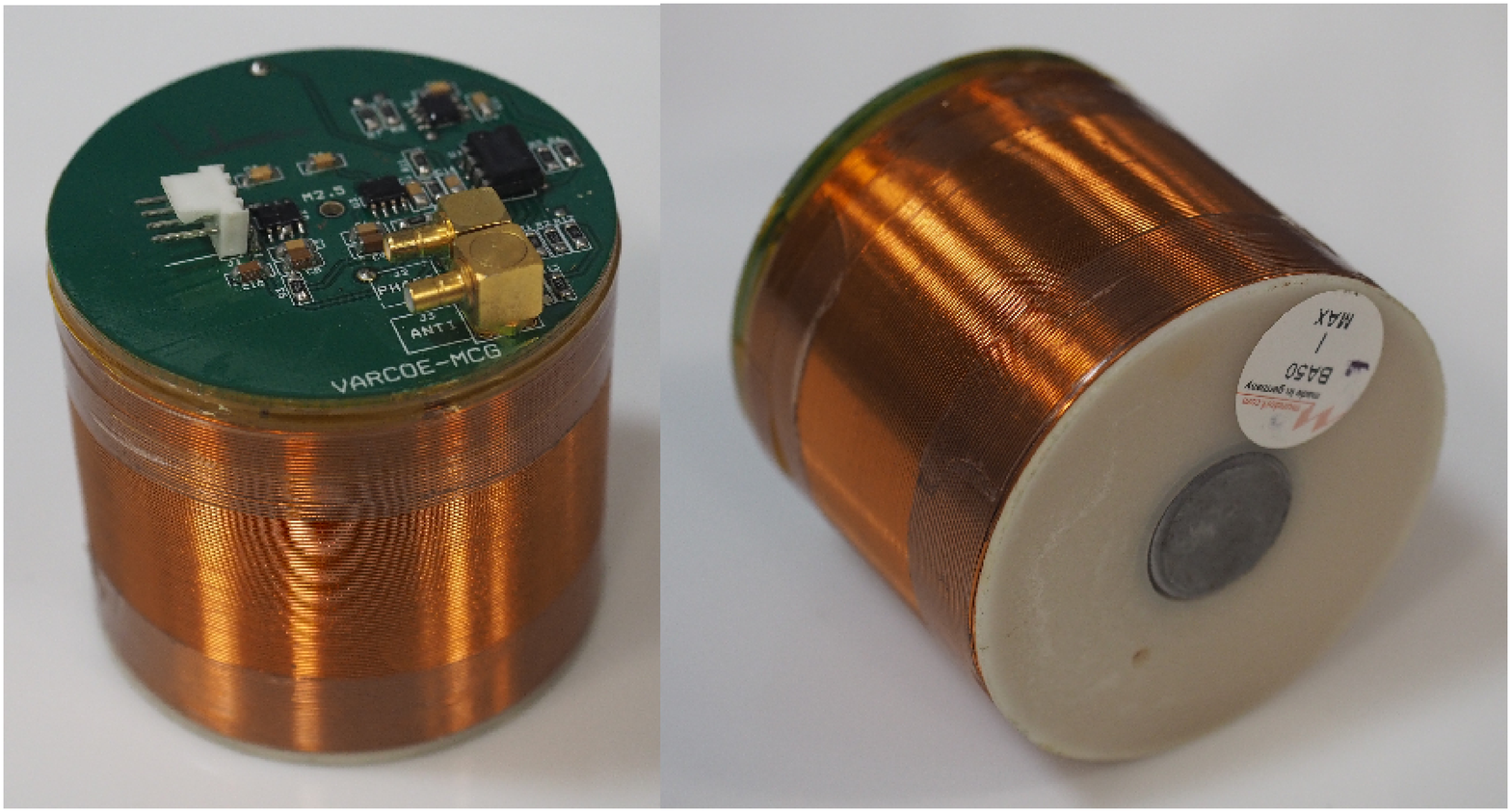}
	\caption{Photograph of the 7cm diameter coil sensors used. The 2cm diameter core is visible in the centre of the coil bobbin. The pre-amplifier circuit board is mounted to the coil to minimise the unshielded signal path. DC battery power is provided via the 4-way header. The amplified signals are output via an SMB connector into a coaxial cable.}\label{CoilSensors}
\end{figure}

\begin{figure}[p]
	\centering
	\includegraphics[width=0.6\textwidth]{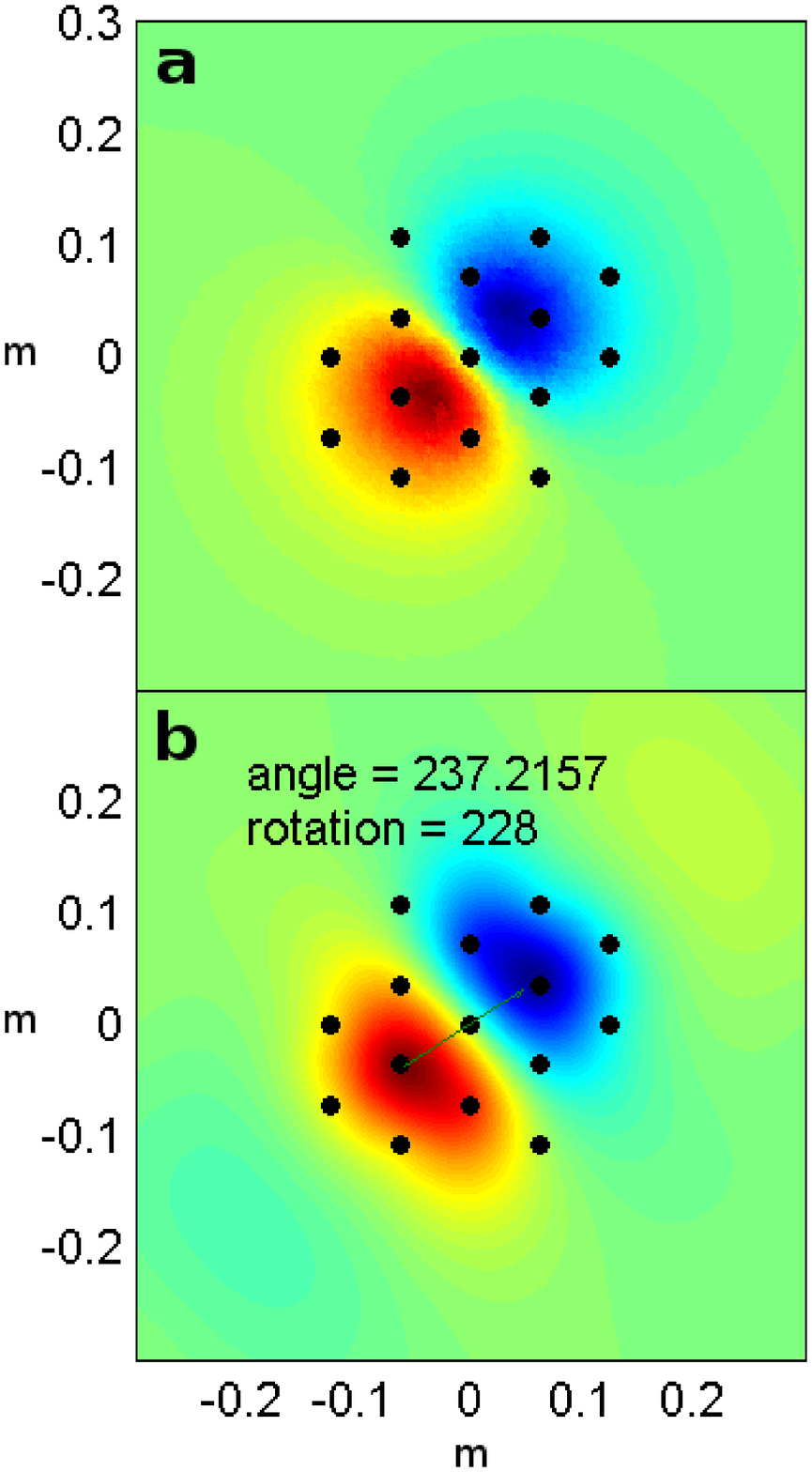}
	\caption{COMSOL model of the sensor array measuring an ideal cardiac dipole field. Figure a shows the cardiac dipole field, which is sampled at the center of the soft iron cores shown as black spots. Figure b shows the interpolated measurement of the field samples. The actual angle and the reconstructed rotation angle are closely matched indicating that an accurate reconstruction is possible with this sensor array.}\label{comsol}
\end{figure}

To evaluate the mapping fidelity of this arrangement a COMSOL model of the array was created. The interaction of the soft iron cores with a static magnetic dipole field comparable in size to the cardiac dipole was simulated. Measurements of the flux at the coil centers were taken as readings equivalent to sensor output. These outputs were spatially interpolated using the same technique as used with actual sensor signals to produce MFM\@. From the MFM the field map angle (FMA) was measured from the vector between the dipole maxima. \Fref{comsol} shows the actual and measured MFM's at a fixed angle.
The simulated cardiac dipole was rotated in small increments. At each increment the difference between the actual angle and the measured angle was calculated. The maximum difference observed was $15^{o}$, with a typical error of $8^{o}$. This uncertainty can be reduced by taking the vector between pole centroids, this spatially averaged measurement of the dipole vector has a considerably lower uncertainty of $<1^{o}$.

\begin{figure}[h]
	\centering
	\includegraphics[width=0.4\textwidth]{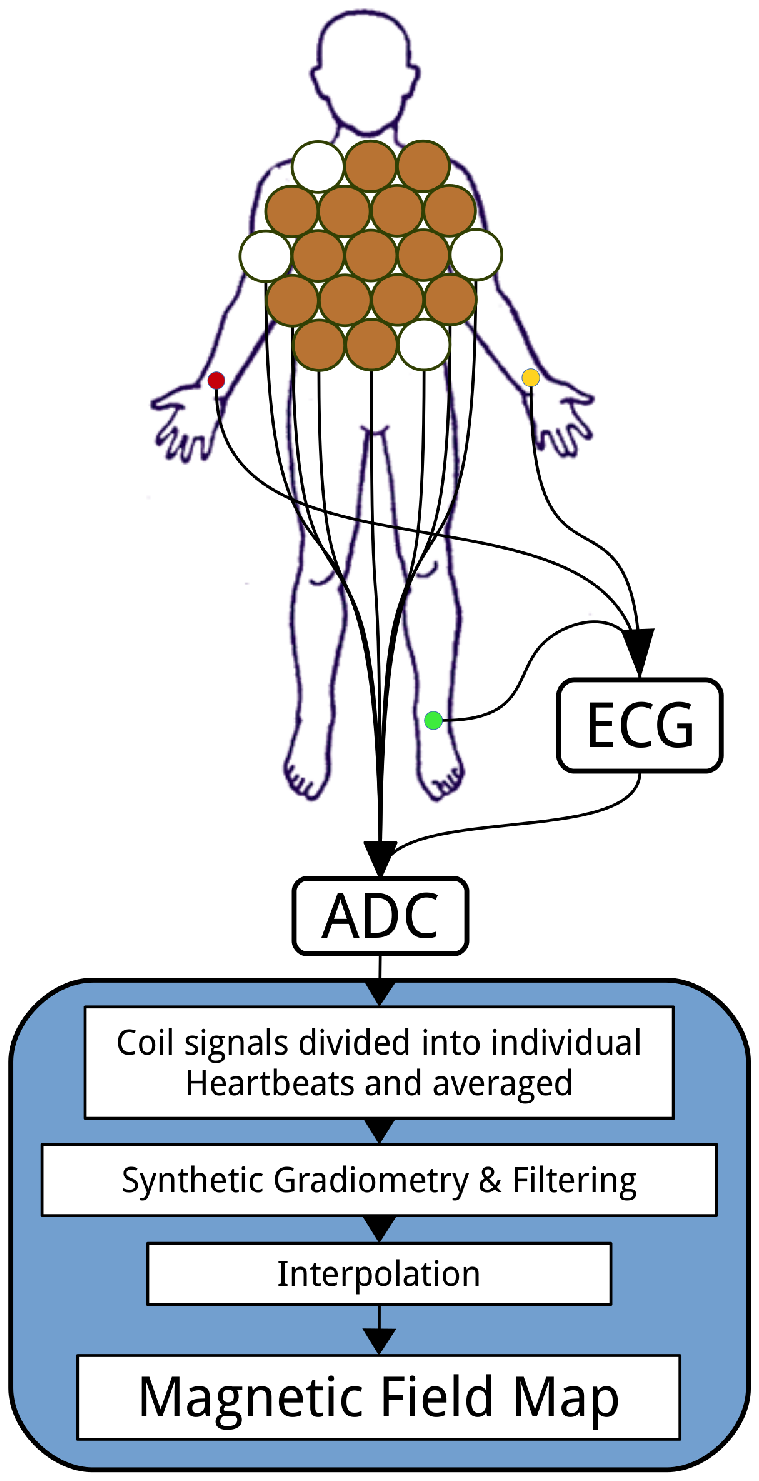}
	\caption{System level diagram; Individual pre-amplified sensor signals are acquired by the ADC, then processed within a computer to create a magnetocardiogram.}\label{flowchart}
\end{figure}

\begin{figure}[h]
	\centering
	\includegraphics[width=0.6\textwidth]{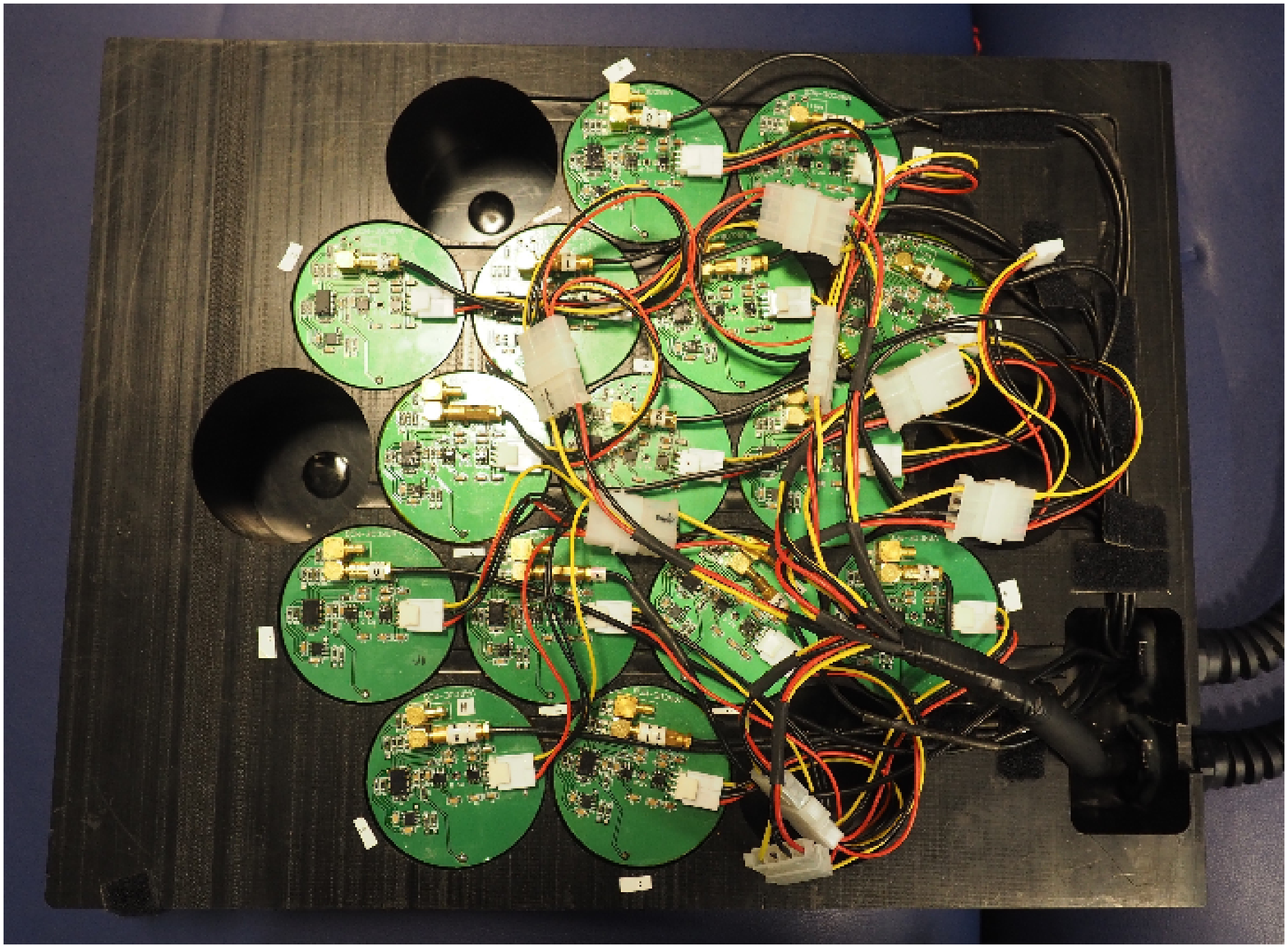}
	\caption{Photograph of the sensor array. The array was machined out of Acetyl engineering plastic, which stiffly couples the sensors together. This array was then bolted to an aluminium frame which supported it above a supine participant in order to capture MCG.}\label{arrayphoto}
\end{figure}

To collect cardiac magnetic fields we designed a mount to hold up to 19 coils with a hexagonal close packed layout of sensor locations, see \Fref{arrayphoto}. The layout, the approximate location against the body and a system level diagram for data acquisition are shown in \fref{flowchart}.
Since movement of the coils within the Earth's field will induce a current in the coils, they must be stiffly coupled so that acoustically induced signals become common mode and therefore removable by gradiometry. To this end, the mount was manufactured from a single piece of Acetal engineering plastic and the coils were securely potted in place. It was supported above the suppine participant by a four legged aluminium frame coupled to the floor. The large mass provided inertial dampening.

\subsection{Data Acquisition and DSP}

\begin{figure}[h]
	\centering
	\includegraphics[width=0.8\textwidth]{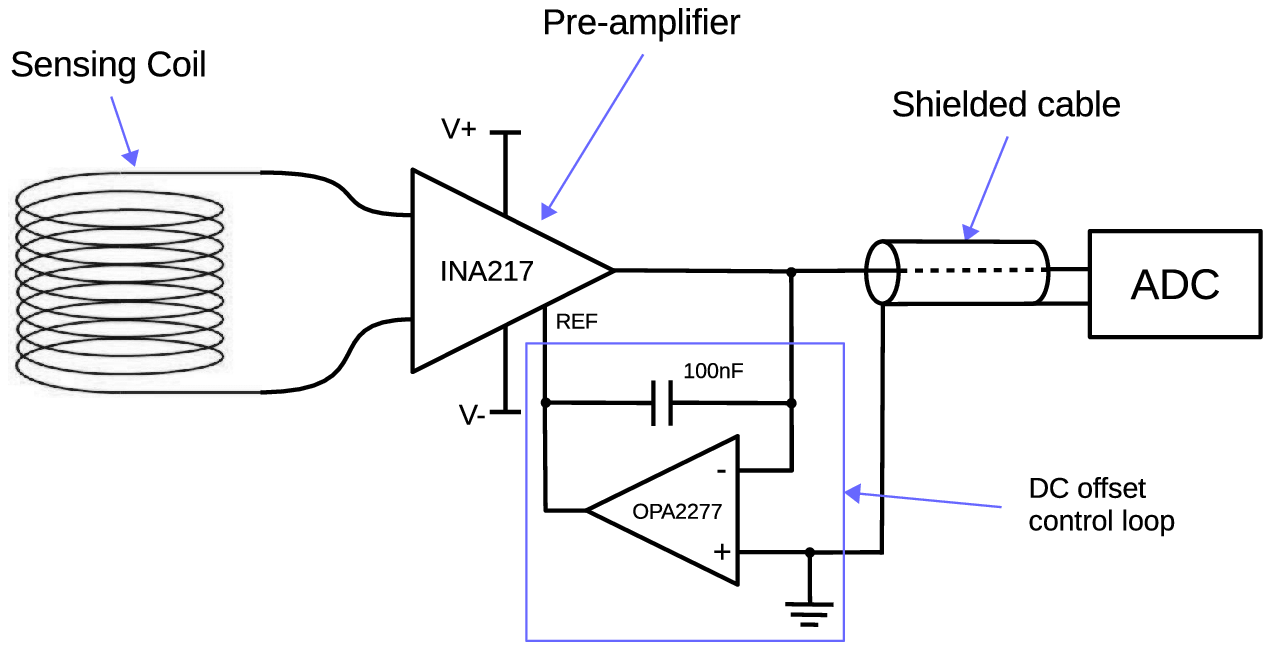}
	\caption{Schematic of a single sensor channel. The DC offset control loop is used to remove drift generated by the amplifier, it functions as a high pass filter with a cut-off of 1.6Hz. The amplifier power supply (V+, V-) is supplied by a pair of batteries with local fixed LDO voltage regulators.}\label{sensorschematic}
\end{figure}

To extract signal from the coils with minimum interference a low noise pre-amplifier ($3.5nV/\sqrt{Hz}$ at 10Hz) with a gain of $1000\times$ was placed immediately above each MFM Coil, see \Fref{sensorschematic} for details.
The signal was then digitized using a National Instruments 16-channel 2kS/s 24bit AC-coupled ADC with a rail-rail voltage of $316mV$ ($37nV$ sensitivity).
Cycle averaging, filtering and gradiometry were performed in digital post processing~\cite{ipython}.

Cycle averaging used the ECG R-wave rising edge as a fiducial. The magnetometer signals were sliced in an interval of $\pm500ms$ about the fiducial, these intervals were then averaged.
A moving average filter was applied by convolution of the signal with a 20ms wide top hat distribution~\cite{numpy}.

In the normal mode of operation of a magnetic gradiometer it is considered necessary to have extensive shielding or closely matched gradiometer coils.
But matching wire wound induction coils to sufficient accuracy is effectively impossible.
However, in an array of coils differences in coil sensitivity are reduced by taking the average, and since the spatial average over a dipole crossection is zero, the cardiac signal is not present in this background.
Hence, this background is a  bucket detector. Subtracting the bucket detector signal from a sensor signal produces a gradiometric signal.

Induction coils measure the time derivative of the magnetic field and not the static field. It is not possible to accurately construct the static field components by integrating their signals since the constant of integration is unknown. Attempting to numerically integrate yeilds signals with large baseline wander. This makes induction coil magnetometers not ideal for low frequency field measurement, though their performance can be improved with a fluxgate arrangement~\cite{fluxgate} or mechanical dithering of the sensors~\cite{MEMs}.
The derivative signals do not contain reliable absolute amplitude information, but they contain the same relative amplitude information and therefore the normalised spatial measurement is unaffected.
Therefore the majority of the diagnostic information is preserved.

\section{Results And Discussion}
\subsection{Sensor Response and Sensitivity}
\begin{figure}[h]
	\centering
	\includegraphics[width=0.8\textwidth]{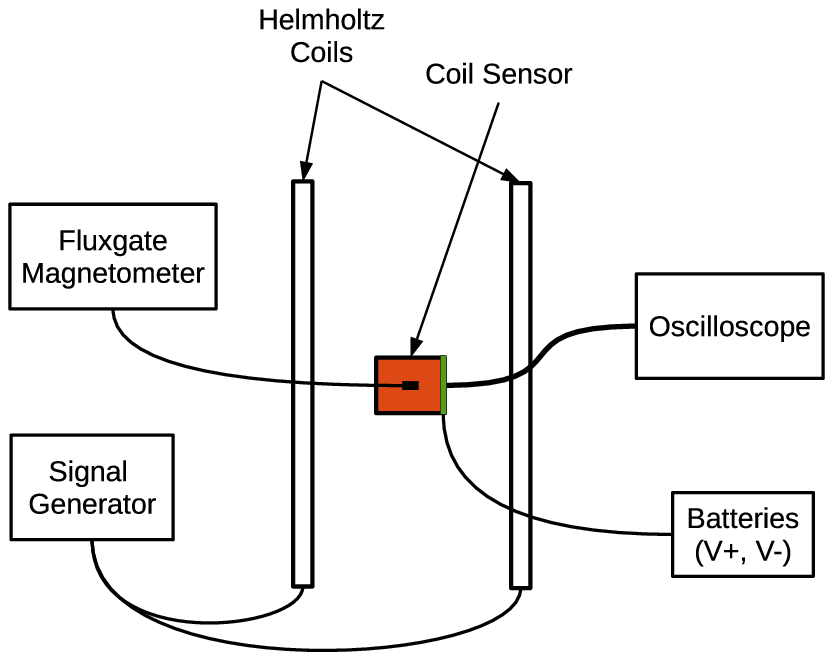}
	\caption{Experiment to measure sensor response. A sinusoidal field was created by the helmholtz coils being driven by the signal generator, the peak field amplitude was measured using a calibrated fluxgate magnetometer. The fluxgate probe was then replaced by a MFM coil sensor and the peak amplitude output by the sensor as recorded using the oscilloscope.}\label{response}
\end{figure}

The sensor response was measured by placing one in the center of a helmholtz coil pair, applying a calibrated sinusoidal field and measuring the sensor output amplitude, as shown in \Fref{response}. The applied magnetic field amplitude was measured using a calibrated fluxgate magnetometer.
The response is dependent on the field frequency and for this coil, it is linear between 1Hz and 1KHz. The measured response was $290fT/\mu V$ at 30Hz and $813fT/\mu V$ at 10Hz.

\begin{figure}[h]
	\centering
	\includegraphics[width=0.8\textwidth]{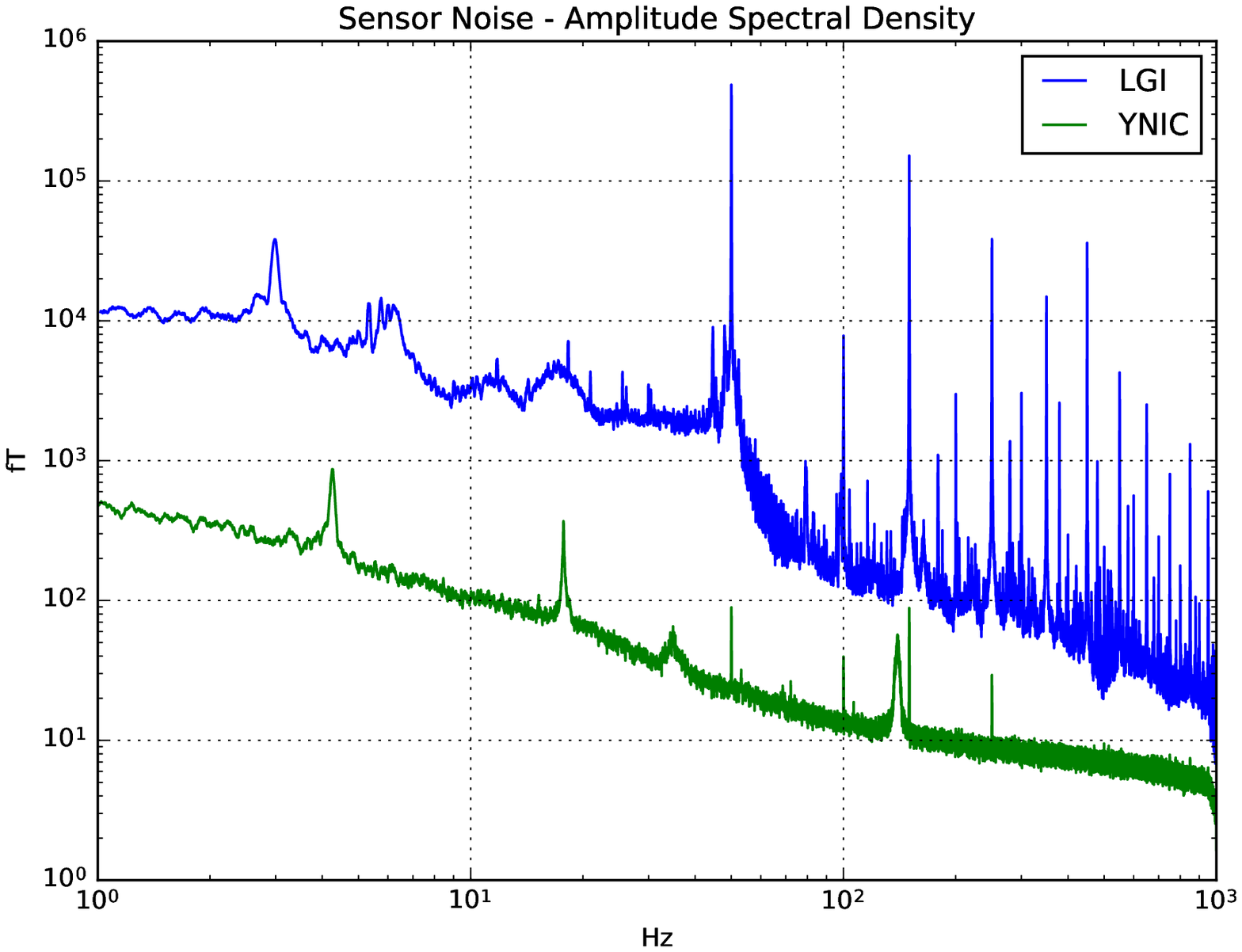}
	\caption{Amplitude spectral density of the sensor in shielded (YNIC) and unshielded (LGI) environments, computed by FFT\@. The sensor voltage spectral density was computed then the frequency dependent response was used to convert to units of magnetic field amplitude (fT).}\label{FFT}
\end{figure}

The sensitivity is determined by the inherent sensor noise. To measure the inherent noise the sensor was placed in a shielded room and 10 minutes of signal were recorded. Computing a FFT on this timeseries gives the amplitude spectral density of the sensor, see \Fref{FFT}. This voltage amplitude spectra was converted into magnetic field amplitude by factoring in the coils frequency response. The resulting noise floor was $104fT/\sqrt{Hz}$ at 10Hz and $36fT/\sqrt{Hz}$ at 30Hz.

\subsection{First MCG Measurements}
\begin{figure}[ht!]
	\centering
	\includegraphics[width=0.8\textwidth]{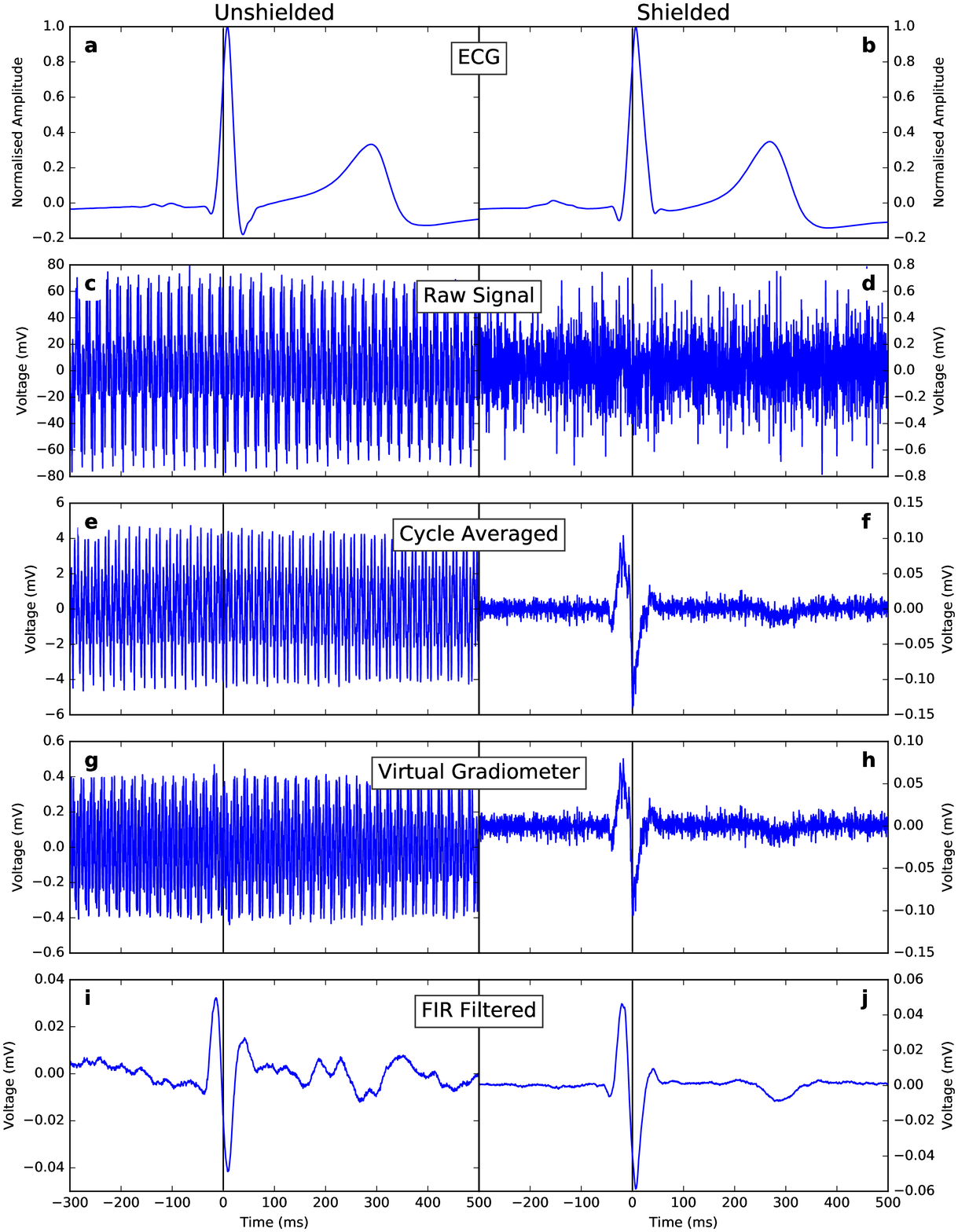}
	\caption{Comparison of two magnetometer signals from the same person taken with an identical device, with similarly positioned sensors but in different environments. The Sheilded MCG was acquired in the York Neuroimaging center (YNIC). The Unsheilded MCG was acquired at Leeds General Infirmary (LGI), 16 months later. The environmental noise is $100\times$ larger in LGI\@. Averaging reduces the noise amplitude by $20\times$. Gradiometry (subtracting the synthetic bucket detector signal) provides $10\times$ noise reduction at LGI but does not effect the noise amplitude at YNIC, in this case it reduces the signal amplitude as the dipole measurement was not symmetric. A final stage FIR filter notches out 50Hz, reducing LGI noise by $500\times$ and acting mostly as a smoothing filter in LGI with $20\times$ reduction. The final signal amplitudes differ by 40\% which could be explained by a difference in sensor positioning or physiological differences. The unshielded signal has a similar level of white noise, but a much larger coloured noise component. }\label{signals}
\end{figure}

Now we show that these coils can detect the magnetic field with sufficient sensitivity and sufficiently low inherent noise. An MCG was taken in a shielded environment, where the environmental noise was low enough to observe dominant sensor noise.
\Fref{signals}d shows a raw MCG sensor signal taken from one of us in a shielded room which had an attenuation of 40dB at 1Hz and a maximum attenuation of 104dB at 30Hz. Aproximately 10 minutes of data were recorded containing 485 cardiac cycles over 12 coils. The noise amplitude was approximately 0.6mV RMS ($<73pT$).

The synchronously recorded 3 lead ECG shown in~\Fref{signals}b was thresholded to find the R-wave rising edge, this was used as a fiducial for cycle averaging, which reduced the noise amplitude by a factor of 35 (to $<2.1pT$). Then the gradient was calculated by subtracting the synthetic bucket detector. However, this did not reduce the noise amplitude since it was composed of Johnson thermal noise which is uncorrelated across the array. Finally a 20ms wide moving average filter was applied to notch out the remaining 50Hz noise and smooth the remaining thermal noise.

The resulting signal has no significant noise ($<150fT$) and corresponds with the anti-derivative of previously observed MCG signals~\cite{koch_reference_2011,kandori_space-time_2008}. It has a maximum amplitude during cardiac depolarisation of 0.05mV ($30pT$).

To analyse the device performance in an unshielded enironment an MCG was recorded at Leeds General Infirmary, the results are shown alongside the shielded data in~\Fref{signals}. Similarly 10 minutes of data were recorded, yeilding 482 cardiac cycles. The noise amplitude is much larger compared to the shielded room signals at 80mV ($\sim20nT$), and highly correlated across the array. Cycle averaging reduces this by a factor of 12. Application of synthetic gradiometry provides $10\times$ rejection. The same 20ms wide moving average filter is highly effective at removing the remaining 50Hz noise and it's harmonics leading to $500\times$ supression.
The resulting signal is the same amplitude as acquired in the shielded room, however there is a large coloured noise content.

This coloured noise could be removed by the aplication of more advanced DSP techniques such as wavelet denoising~\cite{ishikawa_noise_2014}, reference data based non-linear denoising as an alternative improvement to gradiometry~\cite{sternickel_nonlinear_2001}, and EEMD for baseline wander removal~\cite{colominas_2012}. A second layer of coils ontop could provide an improvement to gradiometer performance as the second coils would be coaxial with the first but receive reduced cardiac magnetic flux~\cite{kang_simple_2012}.

The sensitivity to low frequency could be increased by lock-in to a global excitation field provided by a fluxgate or mechanical dithering arrangement~\cite{nakayama_pulse-driven_2011,fluxgate,MEMs,paz_room_2014,jahns_sensitivity_2012}.

\begin{figure}[ht!]
	\centering
	\includegraphics[width=0.8\textwidth]{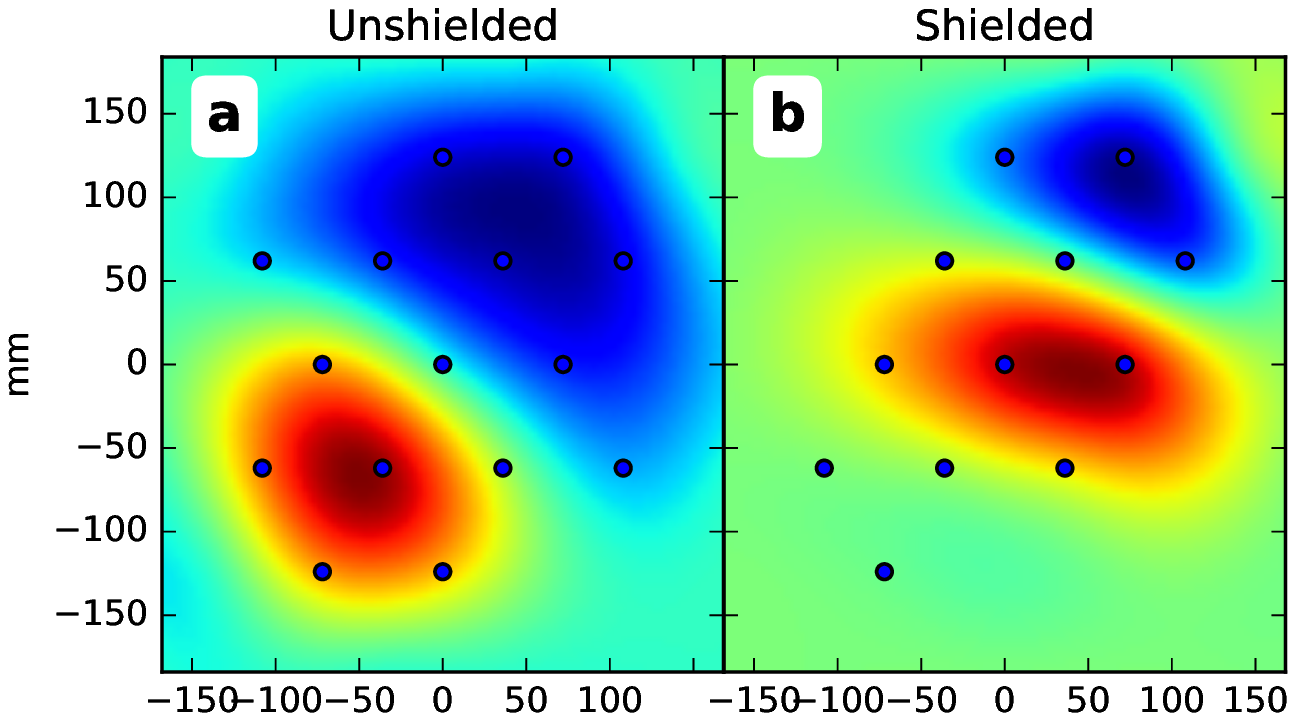}
	\caption{MFM's of the author, during the rising edge of the ECG r-wave. MFM on the left was taken in the YNIC shielded room. The MFM on the right was taken in an unshielded room at Leeds General Infirmary. The shielded measurement used 12 coils as the pre-amplifier wiring broke on three of the coils during transportation of the device to YNIC\@.
	}\label{MFM}
\end{figure}

MFM's represent the magnetic field at a chosen instant in the cardiac cycle. They are created by spatially interpolating the sensor amplitudes from a common time sample.
\Fref{MFM} compares the shielded and unshielded MFM's at -15ms, during the magnetic R wave peak activity.

The dipole angle is consistent between the two MFM's. The dipole position is translated between the MFM's since the array was only subjectively aligned in the coronal plane relative to the Xiphoid process. Also the angle between the coronal and transversal planes was not precisely controlled, since each bed had a different distribution of padding material. Ideally the MFM would be precisely referenced to the individuals cardiac geometry. This would be invaluable for solving the inverse problem; estimating the structure of the underlying current distribution corresponding to the observed MFM\@.

The observed dipole angle and size are in good agreement with past MCG observations of healthy normals~\cite{Lim2009}. We therefore anticipate that the signal has similar diagnostic value.

\section{Conclusion}
We have presented a new design for a device to perform magnetic field mapping and demonstrated that the device collects useful magnetocardiography data in shielded and unshielded environments.

The shielded measurements prove that the coil sensor system has sufficiently low inherent noise for cycle averaged MCG and sufficient spatial resolution for field map angle measurement.

However, operation of the device within an unshielded environment imposes coloured noise on the signal of an amplitude comparable to the repolarisation signals (ECG T wave). This potentially limits the diagnostic capability of our device within unshielded environments. Though the depolarisation signals (QRS) are reliably observable above this noise. This result may be improved by the application of recent developments in denoising algorithms~\cite{ishikawa_noise_2014,sternickel_nonlinear_2001,colominas_2012}.

Further clinical testing will be required to determine if it is capable of detecting recent onset of NSTEMI in patients with the same accuracy as previous devices.
A future device may want to use more sensors to increase the measurement area and may also consider smaller coils to achieve a higher resolution. The addition of a second layer of coils would provide a vertical baseline for synthetic gradiometry which may improve environmental noise suppression~\cite{kang_simple_2012}.
The low frequency performance could be improved to reach DC by lock-in to a global excitation field~\cite{nakayama_pulse-driven_2011,fluxgate,MEMs,paz_room_2014,jahns_sensitivity_2012}.

\section{Acknowledgments}
Permission to use human subjects in the collection of data was granted by the University of Leeds ethics committee (ref.\ no. MEEC 12--034). This research would not have been possible without the assistance of L. Falk, R. Byrom, M. Williamson, D. Brettle, Prof.\ M Kearney and S. Smye at LGI and M. Everitt, S. Brown, L. Burgin, F. Ridgeon, B. Gibbs, P. Thornton, D. Grimmond, M. Moran, and S. Mann at the University of Leeds. We thank Prof.\ G. Green at York Neuroimaging for access to the shielded room. We acknowledge funding from the NHS National Innovation Center, the IKC for Medical Devices, the BHRC, HEIF, IPGroup, Quantum Imaging Limited and the University of Leeds. CS thanks Wellcome Trust and Nuffield for summer project support.

\section*{References}
\bibliographystyle{unsrt}
\bibliography{MFM}

\end{document}